# Galileo and Buonamici on the Tides of the Sea:
# Was Something Omitted from the *Dialogue?*[*]

Christopher M. Graney
Vatican Observatory
00120 Stato Città del Vaticano
c.graney@vaticanobservatory.org

**In his 1616 discourse on the tides, Galileo claimed that diurnal tides (that is, a single high tide and a single low tide each day) occurred in Lisbon, Portugal, bolstering his theory of the tides. Lisbon does not feature such tides, but in an exchange of letters in 1629-1630, Giovanfrancesco Buonamici provided Galileo with information on where such tides could be found. Buonamici referred Galileo to the *Regimiento de Navegación* of Andrés García de Céspedes, and to the *Descrittione di Tutti i Paesi Bassi* of Lodovico Guicciardini. Galileo omitted any information on where diurnal tides occurred from his 1632 *Dialogue*, perhaps unintentionally, leaving him open to criticism. Buonamici's material militates against that criticism.**

## 1 INTRODUCTION

The tides of the sea are a product of the motions Earth undergoes in a heliocentric universe. Those motions should generate a single high tide and a single low tide each day, absent complicating factors. So argued Galileo.

To bolster this argument, Galileo needed to know that such *diurnal* tides were observed to actually occur. He had at one time assumed that they occurred in the Atlantic Ocean, but that assumption turned out to be wrong. In early 1630, Giovanfrancesco Buonamici in Madrid gave Galileo the information he needed regarding the tides: diurnal tides do exist on Earth, just not in the Atlantic as Galileo had assumed.

For some reason, Galileo did not mention this information in the discussion of his tides theory that he included in his 1632 *Dialogue Concerning the Two Chief World Systems*. This omission left his tides theory open to an obvious criticism that the diurnal tides his theory predicted did not exist. And because Galileo had asserted in a 1616 discourse that diurnal tides could be found in the Atlantic, this omission left him open to criticism that he ignored and even suppressed information that contradicted his tidal theory, and indeed that he maintained his theory in the teeth of contrary evidence.

---

[*] **Published in *Journal of Astronomical History and Heritage*, Volume 27, Issue 1: 200 - 208 (2024)**
**https://doi.org/10.3724/SP.J.1440-2807.2024.01.11 | https://articles.adsabs.harvard.edu/full/2024JAHH...27..200G**

Had Galileo simply made reference in the *Dialogue* to the information that Buonamici provided him, those criticisms would have been blunted. His tidal theory would have seemed stronger. Both his contemporary and his modern critics would have had a more difficult task in attacking his ideas regarding the tides. The omission of the Buonamici information seems likely to have been unintentional.

**2 GALILEO'S TIDAL THEORY, IN BRIEF**

The level of the sea is observed to rise and fall every day along seacoasts everywhere. The seas are giant basins filled with water. The seacoasts are the edges of those basins. Galileo argued in 1616 and 1632 (Galilei, 1989: 119-33; Galilei, 2001: 492-508) that there is but one way to make water in a basin rise and fall along the edges of that basin: move the entire basin unevenly. Uneven motion makes the water surge back and forth in the basin; that in turn causes the water at opposite edges of the basin to rise and fall. By contrast, if the basin just sits still, or moves at a steady speed in a straight line, that water just sits in the basin. "If the Earth is immobile," Galileo told Federico Cesi in 1624 (and as he would later state in his *Dialogue*), "the tides are impossible" (Galilei, 1903: 209).[1]

In the heliocentric theory the Earth is not immobile. It rotates diurnally and revolves annually, and those two motions reinforce one another on the side of Earth opposite the sun (the midnight point) and oppose one another on the side of the Earth facing the sun (the noon point). Every point on Earth's surface will be looping around the sun more rapidly at midnight than at noon (Figure 1). The sea basins therefore move unevenly; each day they travel faster, then slower, then faster again (midnight to noon to midnight again), thanks to the combined diurnal and annual motions of the Earth. This, said Galileo, is the uneven motion that drives the tides of the sea.

The driving unevenness is diurnal, so the tides should be diurnal—one high and low tide each day, separated by 12 hours. In the Mediterranean Sea, however, the tides are *semidiurnal*—two high tides and two low tides each day, separated by 6 hours. These semidiurnal tides, Galileo said, arise from the particular characteristics of the Mediterranean (its length and depth) which are such that the surge of water arising from the diurnal driving action of the unevenly-moving surface of Earth reflects back and forth twice daily. Elsewhere on Earth, diurnal tides should be found. Here is how Galileo put this in the discourse he wrote to Cardinal Allesandro Orsini, 8 January 1616 (Galilei, 1989: 127-28):

> I come to resolving the difficulty of how tidal periods can commonly appear to be six hours, even though the primary cause of the tides embodies a principle for moving the water only at twelve-hour intervals, that is, once for the maximum speed of motion and the other for maximum slowness. To this I answer, first, that the determination of the periods which actually occur cannot in any way result from the primary cause alone; instead we must add the secondary cause, depending (as we said) on the specific inclinations of water which, once raised at one end of the vessel [i.e. basin containing the water] by virtue of its own weight, flows to reach equilibrium and undergoes many oscillations and vibrations, more or less frequent depending on the smaller or greater length of the vessel and on the greater or smaller depth of the water. My second answer is that the approximately six-hour period commonly observed is no more natural or significant than any other; rather, it is the one which has been observed and described more than others, since it
takes place in the Mediterranean Sea around which all our ancient
writers and a large part of the moderns have lived. The length of this
Mediterranean basin is the secondary cause that gives its oscillations a
six-hour period; whereas on the eastern shores of the Atlantic Ocean,
which extends to the West Indies, the oscillations have a period of about
twelve hours, as one observes daily in Lisbon, located on the far side of
Spain; now, this sea, which extends toward the Americas as far as the
Gulf of Mexico, is twice as long as the stretch of the Mediterranean from
the Strait of Gibraltar to the shores of Syria, that is, 120 degrees for the
former and 56 degrees for the latter, approximately. Thus, to believe that
tidal periods are six hours is a deceptive opinion and it has led writers to
make up many fictional stories.

> Third, from what has been said it will not be hard to investigate
> the reasons for so many inequalities observed in smaller seas, such as
> the Sea of Marmara and the Hellespont and others.

**3 THE LISBON TIDES PROBLEM**

Galileo believed in 1616 that the Atlantic Ocean at Lisbon, Portugal, was a place where diurnal tides could be found (Shea, 1970: 118; Naylor, 2007: 16). However, this is incorrect. Tides in Lisbon are semidiurnal.

Galileo had been informed of his error by at least 1619. In a letter of 4 April 1619 Tobie Matthew wrote to Francis Bacon, among other things, to relay information from Richard White, who had been in Florence and had spoken with Galileo (Bacon, 1871: 211):[2]

> He [White] tells me, that Galileo had answered your discourse
> concerning the flux and reflux [i.e. tides] of the sea, and was sending it
> unto me; but that Mr. White hindered him, because his answer was
> grounded upon a false supposition, namely, that there was in the ocean
> a full sea [i.e. high tide] but once in twenty-four hours.

A decade later, as Galileo worked on completing his *Dialogue*, he turned to Buonamici in Madrid for information on the seas. In a letter of 19 November 1629, Galileo wrote, "Your Lordship should know that I am about to finish some dialogues in which I treat of the constitution of the universe, and among the principal problems I write of the ebb and flow of the sea, and I believe I have found the true reason for it.... I estimate it to be true and so do all of those with whom I have conferred about it". And Galileo asked if Buonamici would tell him more about what those who have travelled the seas have to say about them (Fantoli, 2003: 241-242).

In the letter, Galileo specifically mentioned currents and tides in the Straits of Magellan and of Gibraltar, and the strait between the island of San Lorenzo (Madagascar) and the African coast. He mentioned the sea at the lighthouse at Messina. And he said that, in brief, "the more particulars I could know, the more helpful they would be to me, because the histories of things sensible are the principles upon which are established the sciences".[3]

Buonamici (1904: 74) responded in a letter of 1 February 1630. "I do not recall having seen anyone who discusses the tides of the sea better than Lodovico Guicciardini in the *Descrittione di Tutti i*

*Paesi Bassi* [Figure 2] in the chapter on the sea," he wrote. He also relayed information that Galileo was no doubt happy to hear: Andrés García de Céspedes in his *Regimiento de Navegación* (Figure 3) states that in the East Indies there are places where the tide is diurnal.[4]

Buonamici gave Galileo good information. In Chapter 33 of *Regimiento*, "Which treats on the waxing and waning of the sea", Céspedes (1606: 79) indeed cites "la India Oriental", where "the sea does not wax nor wane more than once in 24 hours".[5] Buonamici cautioned Galileo that Céspedes does not specify exact locations for this diurnal tide, nor does he cite specific authorities. Buonamici was again correct—Céspedes indeed does not. Still, Céspedes was a source that said that there were places diurnal tides existed; not Lisbon, but the East Indies.

Céspedes was also correct. Tides occur semidiurnally throughout most of the world, but there are broad regions where tidal behavior is more complex. There are even regions where tides are in fact diurnal. These include areas in Indonesia, what was once commonly called the East Indies, which notably feature diurnal tides, as well as areas of the Gulf of Mexico and the Caribbean, what was once commonly called the West Indies.[6] The current science regarding this tidal behavior is not Galileo's, of course, but the behavior Galileo was looking for does exist, and Buonamici had found good information about it for him.

Buonamici gave Galileo good information regarding Guicciardini as well as Céspedes. However, Galileo would have found the information in *Descrittione* to have been less encouraging than what Céspedes had provided. Guicciardini (1588: 26-27) discusses tides occurring semidiurnally in many places on the Atlantic coast: Ireland, Spain, even far south in Africa. Then he adds, "and this applies to the Ocean in general, leaving aside the East and West Indies".[7]

**4 DIURNAL TIDES IN THE *DIALOGUE***

In the *Dialogue*, which Galileo submitted for publication in Rome in the summer of 1630, he wrote almost the same thing regarding tidal periods as he wrote in the discourse for Orsini in 1616, except that he omitted any mention of the Atlantic Ocean, Lisbon, and the 120-degree span of the Atlantic versus the shorter span of the Mediterranean. Here are both discussions, side-by-side for comparison:

| **Orsini (1616)** | ***Dialogue* (1632)** |
|---|---|
| I come to resolving the difficulty of how tidal periods can commonly appear to be six hours, even though the primary cause of the tides embodies a principle for moving the water only at twelve-hour intervals, that is, once for the maximum speed of motion and the other for maximum slowness. To this I answer, first, that the determination of the periods which actually occur cannot in any way result from the primary cause alone; instead we must add the secondary cause, depending (as we said) on the | Now, secondly, I shall resolve the question why, since there resides in the primary principle no cause of moving the waters except from one twelve-hour period to another (that is, once by the maximum speed of motion and once by the maximum slowness), the period of ebbing and flowing nevertheless commonly appears to be from one six-hour period to another. Such a determination, I say, can in no way come from the primary cause alone. The secondary causes must be |

specific inclinations of water which, once raised at one end of the vessel by virtue of its own weight, flows to reach equilibrium and undergoes many oscillations and vibrations, more or less frequent depending on the smaller or greater length of the vessel and on the greater or smaller depth of the water. My second answer is that the approximately six-hour period commonly observed is no more natural or significant than any other; rather, it is the one which has been observed and described more than others, since it takes place in the Mediterranean Sea around which all our ancient writers and a large part of the moderns have lived. The length of this Mediterranean basin is the secondary cause that gives its oscillations a six-hour period; whereas on the eastern shores of the Atlantic Ocean, which extends to the West Indies, the oscillations have a period of about twelve hours, as one observes daily in Lisbon, located on the far side of Spain; now, this sea, which extends toward the Americas as far as the Gulf of Mexico, is twice as long as the stretch of the Mediterranean from the Strait of Gibraltar to the shores of Syria, that is, 120 degrees for the former and 56 degrees for the latter, approximately. Thus, to believe that tidal periods are six hours is a deceptive opinion and it has led writers to make up many fictional stories.

      Third, from what has been said it will not be hard to investigate the reasons for so many inequalities observed in smaller seas, such as the Sea of Marmara and the Hellespont and others.

introduced for it; that is, the greater or lesser length of the vessels and the greater or lesser depth of the waters contained in them. These causes, although they do not operate to move the waters (that action being from the primary cause alone, without which there would be no tides), are nevertheless the principal factors in limiting the duration of the reciprocations, and operate so powerfully that the primary cause must bow to them. Six hours, then, is not a more proper or natural period for these reciprocations than any other interval of time, though perhaps it has been the one most generally observed because it is that of our Mediterranean, which has been the only place practicable for making observations over many centuries. Even so, this period is not observed everywhere in it; in some of the narrower places, such as the Hellespont and the Aegean, the periods are much briefer. (Galilei, 2001: 502).

The *Dialogue* discussion parallels the Orsini discussion, but is shorter, owing largely to the omission of the material on the Atlantic and Lisbon. Note that in the *Dialogue*, Galileo simply states that the semidiurnal tides were not "more proper or natural"; rather, they were just what has been "most generally observed" because that is what is seen in the Mediterranean, "the only place practicable for making observations over many centuries." And even there, variations exist—"in some of the narrower places, such as the Hellespont and the Aegean, the periods are much briefer". But he provides no example of where diurnal tides can be found, even though Buonamici had told him that they can be found in the East Indies.

Some modern scholars, not fully considering Galileo's exchange of letters with Buonamici, have suggested that Galileo's retention in the *Dialogue* of his general explanation of the semidiurnal tides is an example of a scientist clinging to his theory in the face of clearly contrary data. After all, what Galileo had said in 1616 about how much the particular characteristics of the Atlantic differed from those of the Mediterranean was entirely correct, and yet both seas experience semidiurnal tides, whereas according to his theory their tidal periods should differ (Shea, 1970: 118; Palmerino, 2004: 218—in agreement with Shea; Hooper, 2004: 228; Graney, 2019: 87, 97; Naylor, 2001: 347). Ron Naylor has stated that "no amount of adverse evidence seemed capable of shaking [Galileo's] resolve" regarding his tides theory, and that he was "too personally involved in the project to be capable of seeing the issue of the tidal theory objectively [2007: 347]". William Shea and Mariano Artigas have argued that "[Galileo] was so convinced of the validity of his proof of the Earth's motion that he continued to believe, in the teeth of evidence, that the diurnal period in the ocean followed a 24- and not a 12-hour cycle. His faith in his theory was greater than his trust in what sailors reported [132]."[8]

Galileo did not merely cling to his theory in the teeth of contrary evidence. He omitted from the *Dialogue* the contrary evidence of Lisbon and the Atlantic, evidence of which he was of course aware. That move would seriously damage the reputation, credibility and career of any scientist. Or so all this would seem, not considering Galileo's exchange of letters with Buonamici.

Buonamici's letter shows that Galileo in 1630 still had evidence, in the form of the word of an authority (Céspedes), that there were places on Earth where the ocean revealed a 24-hour cycle, exactly as he had once imagined at Lisbon. Galileo was wrong about Lisbon. He was wrong about "the Ocean in general, leaving aside the East and West Indies" (to borrow Guicciardini's phrasing). But he was not completely wrong. He easily could have substituted the East Indies for Lisbon in the *Dialogue*'s tides discussion. Indeed, it is difficult to read the *Dialogue* tides discussion in light of Buonamici's letter without suspecting an error of omission.

**5 THE CASE FOR AN ERROR OF OMISSION**

This is not to say that what Galileo learned from Buonamici would not have required some re-evaluation of Galileo's explanation for differing tidal periods. Guicciardini's note that semidiurnal tides are what are found in the ocean generally, with other behavior being limited to the East and West Indies, would certainly reinforce what Galileo had learned in 1619 about Lisbon: seas with widely differing characteristics often share the same tidal periods. If the histories of things sensible are the principles upon which are established the sciences, then the story told by Guicciardini strongly challenged the theory that semidiurnal tides arose from the particular characteristics of different seas.

Or perhaps not. Galileo's explanation for semidiurnal tides was already speculative. He did not know the depths of the Mediterranean and Atlantic. It would be easy enough for him to speculate further, about some general proportion existing between breadth and depth in seas, perhaps—a proportion that manifested itself in the semidiurnal tides that are found in many places, just like the more complex tidal periods found in places like the Hellespont revealed the complex structure of the sea basins in those locations.

The fact remained that if the Earth was immobile, it was impossible to explain the tides without invoking ideas for which there was no solid physics (Newton's ideas being well in the future). Consider what Galileo said in the *Dialogue* regarding Kepler's idea about the moon and the tides:

> But among all the great men who have philosophized about this
> remarkable effect, I am more astonished at Kepler than at any other.
> Despite his open and acute mind, and though he has at his fingertips the
> motions attributed to the earth, he has nevertheless lent his ear and his
> assent to the moon's dominion over the waters, to occult properties, and
> to such puerilities. (Galilei, 2001: 536)

Galileo had a non-occult mechanism to explain the tides. He had information indicating that the diurnal tides that his mechanism required did indeed exist. The rest was speculative details, and could be addressed in a few sentences in the *Dialogue*, replacing those sentences about the Atlantic tides that were in the discourse to Orsini. Such sentences would squelch critics. The absence from the *Dialogue* of any replacement for the Atlantic and Lisbon sentences in the Orsini discourse is puzzling; one wonders if this absence was unintentional.

Consider Zaccaria Pasqualigo's criticism of the *Dialogue*'s tides discussion (Finocchiaro, 1989: 35, 274). Pasqualigo was one of the three members of the special commission appointed by Pope Urban VIII to investigate the *Dialogue*. He wrote the following regarding Galileo's ideas about the tides:

> However, he does not untangle the difficulty that, given this doctrine,
> since the change between greatest acceleration and maximum
> retardation of the earth's motion occurs at twelve-hour intervals, then
> high and low tides should also occur at twelve-hour intervals. But
> experience teaches that they occur every six hours.

This is a valid and easy criticism, given the *Dialogue* as it is, lacking any reference to diurnal tides existing anywhere. (Presumably Pasqualigo was unaware of the 1616 discourse, which would have given him grounds for further criticism, of a more cutting nature, regarding Galileo and the Atlantic tides.) But had Galileo included in the *Dialogue* the information from Céspedes about where tides with a twelve-hour interval could be found, and a bit more speculation on ocean basins, Pasqualigo's criticism would have been largely stymied.

**6 CONCLUSION**

Galileo's argument that the tides of the sea are a product of the motions of the Earth in a heliocentric universe needed diurnal tides to bolster it, because the driving action resulting from those motions would be diurnal. If diurnal tides existed, he could explain away other tidal periods as being a result of the local characteristics of sea basins. Given that when writing the *Dialogue* Galileo had on hand (thanks to Giovanfrancesco Buonamici) information from Andrés García de Céspedes on diurnal tides occurring in the East Indies, and given the reduced length of the *Dialogue*'s discussion of tidal periods and its content compared to Galileo's 1616 discourse on tides to Cardinal Orsini—which included (in error) mention of diurnal tides occurring in the Atlantic at Lisbon—it seems reasonable to suppose that Galileo somehow overlooked adding the Céspedes information to the *Dialogue*. It seems as reasonable to make this supposition as it is to suppose (as has been done in the absence of awareness of Buonamici's work) that Galileo clung to an idea in the teeth of adverse evidence, evidence that he even suppressed, leaving himself open to criticisms of his tidal theory in his time, and to the criticisms of modern scholars today.

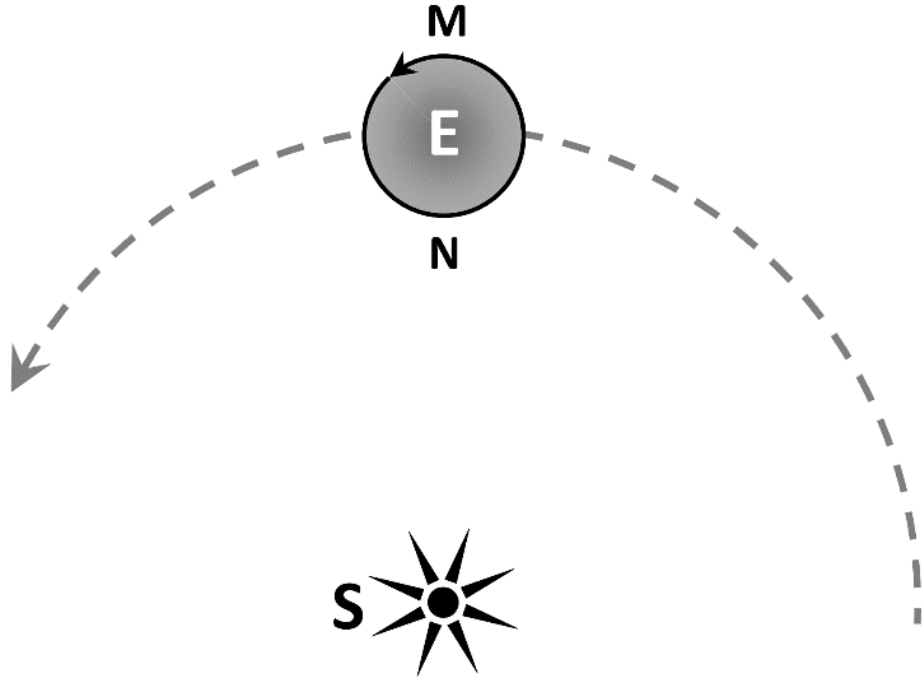

FIGURE 1

Earth E revolves in its orbit counter-clockwise around the sun S, while rotating counter-clockwise around its own axis. At the midnight point M the two motions add together, whereas at the noon point N the rotational motion subtracts from the orbital. Thus at M a point on the surface of the Earth is looping around the sun more rapidly than at N.

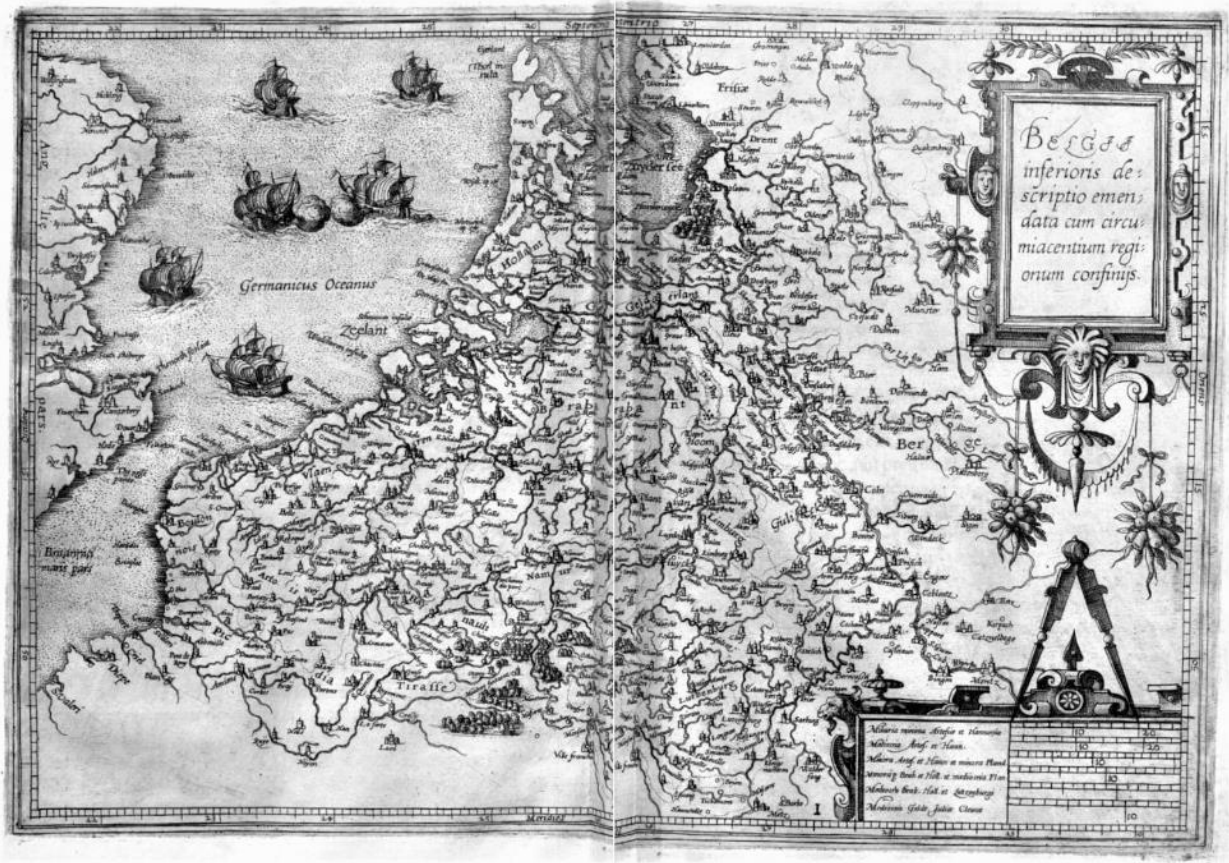

FIGURE 2

Seacoast along the English Channel, from Guicciardini's *Descrittione* (image credit: Google Books).

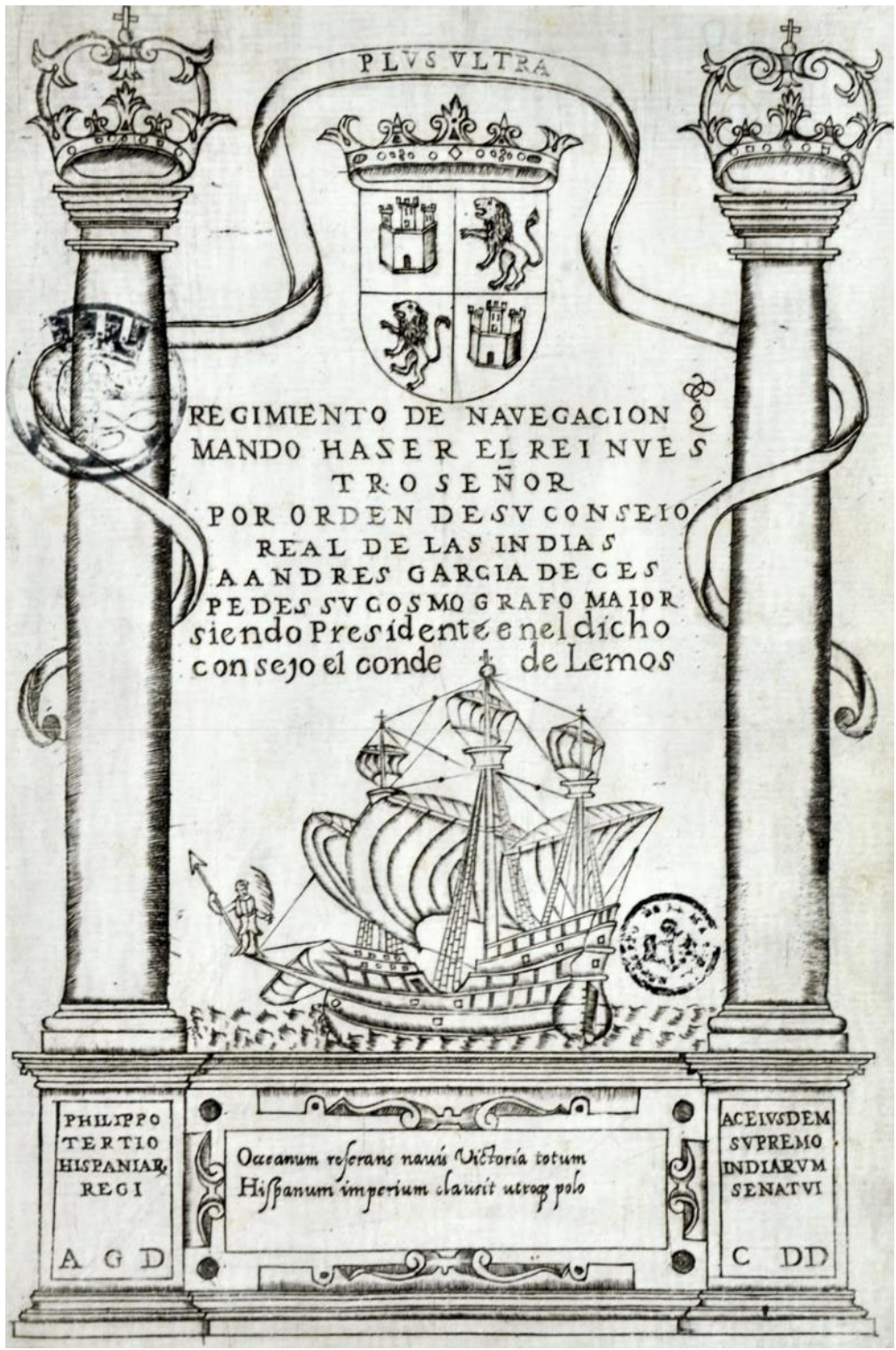

FIGURE 3 (optional)

Title page of *Regimiento de Navegación* (image credit: Google Arts & Culture/Museo Naval, Madrid).

[1] "Stando la terra immobile, è impossibile che seguano i flussi e reflussi." See also Shea and Artigas (2003), 121; Omodeo (2014), 314; Galilei (2001), 484. For thorough general discussions of Galileo's tidal theory, see Naylor (2007); Palmieri (1998).

[2] See also Drake (1978/2003), 273-74; Hooper (2004), 227-28; Graney (2019), 86. For recent discussion of Matthew, Bacon, White and Galileo, see Giudice (2020).

[3] "deve V. S. sapere come sono sul finire alcuni Dialogi ne i quali tratto la costituzione dell'universo, e tra i problemi principali scrivo del flusso e reflusso del mare, dandomi a credere d'haverne trovata la vera cagione.... Io la stimo vera, e tale la stimano tutti quelli con i quali io l'ho conferita."; "et in somma quanti più particolari io potessi sapere, più mi sarebbono grati, perchè l'istorie, cioè le cose sensate, sono i principii sopra i quali si stabiliscono le scienze" —Galilei (1904), 54.

[4] "Circa il flusso et reflusso, io non mi ricordo haver visto nessuno che ne discorra meglio di Lodovico Guicciardini nella Descritione de' Paesi Bassi, nel capitolo del mare". See Céspedes (1606), 75. David Wootton (2010) mentions this (p. 192), but gives few details.

[5] "Que trata de las crecientes, menguantes de la mar"; "donde no crece ni mengua la mar mas de vna vez en 24. horas."

[6] See Ray, Egbert, and Erofeeva (2005), which notes how "diurnal tides are unusually strong and are dominant along some coastlines" in the area of Indonesia (p. 74). Also see National Ocean Service. Areas in Indonesia are among those where tides are diurnal. Areas of the Gulf of Mexico and the Caribbean also have diurnal tides and other complex tidal behavior known as "mixed semidiurnal" tides.

[7] "Et questo basti quanto al generale dell'Oceano, lasciando di parlare de Mari Indiani Orientali, & Occidentali."

[8] Naylor mentions Galileo's letter to Buonamici (citing the Fantoli reference above) but not Buonamici's letter to Galileo. Shea and Artigas briefly mention the 1629-1630 exchange of letters between Galileo and Buonamici but state that "Buonamici made enquiries and confirmed that the flow and ebb of the sea followed a 12- and not a 24-hour cycle"; this assessment seems to conflict with Buonamici's mention of Céspedes. Wootton (2010: 296) states that the exchange between Galileo and Buonamici is "misrepresented" by Artigas and Shea. For a dramatic example of a scholar negatively assessing Galileo in this regard, see Blåsjö (2021).